\documentclass[english, amsmath, amssymb, 12pt, superscriptaddress, tightenlines]{revtex4}
\usepackage{lmodern}
\usepackage[T1]{fontenc}
\usepackage[latin9]{inputenc}
\usepackage{amsmath}
\usepackage{graphicx}
\setcitestyle{round}

\makeatletter
\@ifundefined{textcolor}{}
{%
 \definecolor{BLACK}{gray}{0}
 \definecolor{WHITE}{gray}{1}
 \definecolor{RED}{rgb}{1,0,0}
 \definecolor{GREEN}{rgb}{0,1,0}
 \definecolor{BLUE}{rgb}{0,0,1}
 \definecolor{CYAN}{cmyk}{1,0,0,0}
 \definecolor{MAGENTA}{cmyk}{0,1,0,0}
 \definecolor{YELLOW}{cmyk}{0,0,1,0}
}
\numberwithin{equation}{section}
\numberwithin{figure}{section}

\makeatletter
\newcommand*{\citenumns}[2][]{%
  \begingroup
  \let\NAT@mbox=\mbox
  \let\@cite\NAT@citenum
  \let\NAT@space\NAT@spacechar
  \let\NAT@super@kern\relax
  \renewcommand\NAT@open{}%
  \renewcommand\NAT@close{}%
  \cite[#1]{#2}%
  \endgroup
}
\makeatother

\usepackage{upgreek}
\usepackage{babel}
\usepackage{lmodern}
\usepackage[T1]{fontenc}
\usepackage[latin9]{inputenc}
\usepackage[pdftex]{color}
\usepackage{amstext}
\usepackage{amssymb}

\usepackage{chngcntr}
\counterwithout{figure}{section}

\newcommand{\BYOO}{Ba$_2$YOsO$_6$}

\newcommand{\CLOO}{Ca$_3$LiOsO$_6$}

\makeatother

\begin{document}


\title{Title:  Spin-orbit coupling controlled $J=3/2$ electronic ground state in
5$d^{3}$ oxides }

\author{\textbf{Authors:}  A. E. Taylor }

\address{Quantum Condensed Matter Division, Oak Ridge National Laboratory,
Oak Ridge, Tennessee 37831, USA}

\author{S. Calder}

\address{Quantum Condensed Matter Division, Oak Ridge National Laboratory,
Oak Ridge, Tennessee 37831, USA}

\author{R. Morrow}

\address{Department of Chemistry, The Ohio State University, Columbus, Ohio
43210-1185, USA}

\author{H. L. Feng}

\address{Research Center for Functional Materials, National Institute for
Materials Science, 1-1 Namiki, Tsukuba, Ibaraki 305-0044, Japan}

\author{M. H. Upton}

\address{Advanced Photon Source, Argonne National Laboratory, Argonne, Illinois
60439, USA}

\author{M. D. Lumsden}

\address{Quantum Condensed Matter Division, Oak Ridge National Laboratory,
Oak Ridge, Tennessee 37831, USA}

\author{K. Yamaura }

\address{Research Center for Functional Materials, National Institute for
Materials Science, 1-1 Namiki, Tsukuba, Ibaraki 305-0044, Japan}

\author{P. M. Woodward}

\address{Department of Chemistry, The Ohio State University, Columbus, Ohio
43210-1185, USA}

\author{A. D. Christianson{*}}

\address{Quantum Condensed Matter Division, Oak Ridge National Laboratory,
Oak Ridge, Tennessee 37831, USA}

\address{Department of Physics and Astronomy, The University of Tennessee,
Knoxville, TN 37996, USA \\ 
*To whom correspondence should be addressed:  Email: christiansad@ornl.gov}

\baselineskip24pt

\maketitle

\noindent
 \textbf{Abstract:  Entanglement of spin and orbital degrees of freedom drives the formation of novel quantum and topological physical states. Discovering new spin-orbit entangled ground states and emergent phases of matter requires both experimentally probing the relevant energy scales and applying suitable theoretical models. Here we report resonant inelastic x-ray scattering measurements of the transition metal oxides Ca$_3$LiOsO$_6$ and Ba$_2$YOsO$_6$. We invoke an intermediate coupling approach that incorporates both spin-orbit coupling and electron-electron interactions on an even footing and reveal the ground state of $5d^3$ based compounds, which has remained elusive in previously applied models, is a novel spin-orbit entangled J=3/2 electronic ground state. This work reveals the hidden diversity of spin-orbit controlled ground states in 5d systems and introduces a new arena in the search for spin-orbit controlled phases of matter.}


\noindent
\\
\textbf{Main Text:}  The electronic ground state adopted by an ion is a fundamental determinant of manifested physical properties. Recently, the importance of spin-orbit coupling (SOC) in creating the electronic ground state in 5$d$-based compounds has come to the fore and revealed novel routes to a host of unconventional physical states including quantum spin liquids, Weyl semimetals, and axion insulators~\cite{jackeli_mott_2009,witczak-krempa_correlated_2014}.  As a result, major experimental and theoretical efforts have been
undertaken seeking novel spin-orbit physics in various $5d$ systems, and the influence of SOC has now been observed in the macroscopic 
properties of numerous systems. However, beyond the $J_{\mathrm{eff}}=1/2$ case such as that found in Sr$_2$IrO$_4$~\cite{kim_phase-sensitive_2009}  --- which is a single-hole state that applies only to idealised 5$d^5$ ions in cubic materials --- questions abound concerning the electronic ground states which govern $5d$ ion interactions.

In this context $5d^{3}$  materials present a particularly intriguing puzzle, because octahedral $d^{3}$ configurations are expected to be orbitally-quenched $S=3/2$ states --- in which case
SOC enters only as a 3rd order perturbation~\cite{sugano_multiplets_1970} --- yet there is
clear experimental evidence that SOC influences the magnetic
properties in 5$d^{3}$ transition metal oxides (TMOs). This includes the observations of large, SOC-induced spin-gaps in their magnetic excitation spectra~\cite{kermarrec_frustrated_2015,taylor_spin-orbit_2016,calder_spin-orbit-driven_2016} and x-ray absorption branching ratios which deviate from $\mathrm{BR} = I_{\mathrm{L}3}/I_{\mathrm{L}2} =2$~\cite{laguna-marco_electronic_2015,veiga_fragility_2015}. Despite this, no description beyond the $S=3/2$ state had been established. The emergent phenomena in 4$d^{3}$ and 5$d^{3}$ systems, such as
incredibly high magnetic transition temperatures~\cite{shi_crystal_2010,thorogood_structural_2011,rodriguez_high_2011,krockenberger_sr2croso6:_2007},
Slater insulators~\cite{calder_magnetically_2012}, and possible
Mott-insulators~\cite{meetei_theory_2013} are therefore poorly understood. 

We selected $5d^{3}$ TMOS \CLOO{} and \BYOO{} as model systems in which to investigate the influence of spin-orbit coupling on the electronic ground states via RIXS
measurements on polycrystalline samples. Both materials have relatively-high
magnetic ordering temperatures, $T_{\mathrm{N}}=117$ and 67\,K,
respectively, despite large separation of nearest-neighbour Os
ions of 5.4--5.9~$\mathrm{\AA}$~\cite{kermarrec_frustrated_2015,shi_crystal_2010,calder_magnetic_2012}. The
relative isolation of Os-O octahedra allows us to unambiguously
access the ground state, because only extended superexchange
interactions are present. Stronger, close-range interactions can mask the effective single-ion
levels we wish to observe~\cite{ament_resonant_2011}.  In \CLOO{} the oxygen
octahedra surrounding Os$^{5+}$ ions are very close to ideal, despite
the overall non-cubic symmetry --- hexagonal \emph{R$\bar{3}$c}~\cite{shi_crystal_2010}.
As previously reported~\cite{kermarrec_frustrated_2015}, we find that \BYOO{} has the ideal cubic double perovskite structure \emph{Fm$\bar{{3}}$m}
to within experimental measurement limits, see supplementary material
for high-resolution synchrotron x-ray and neutron diffraction.

\begin{figure}[t!]
\includegraphics[width=0.5\columnwidth]{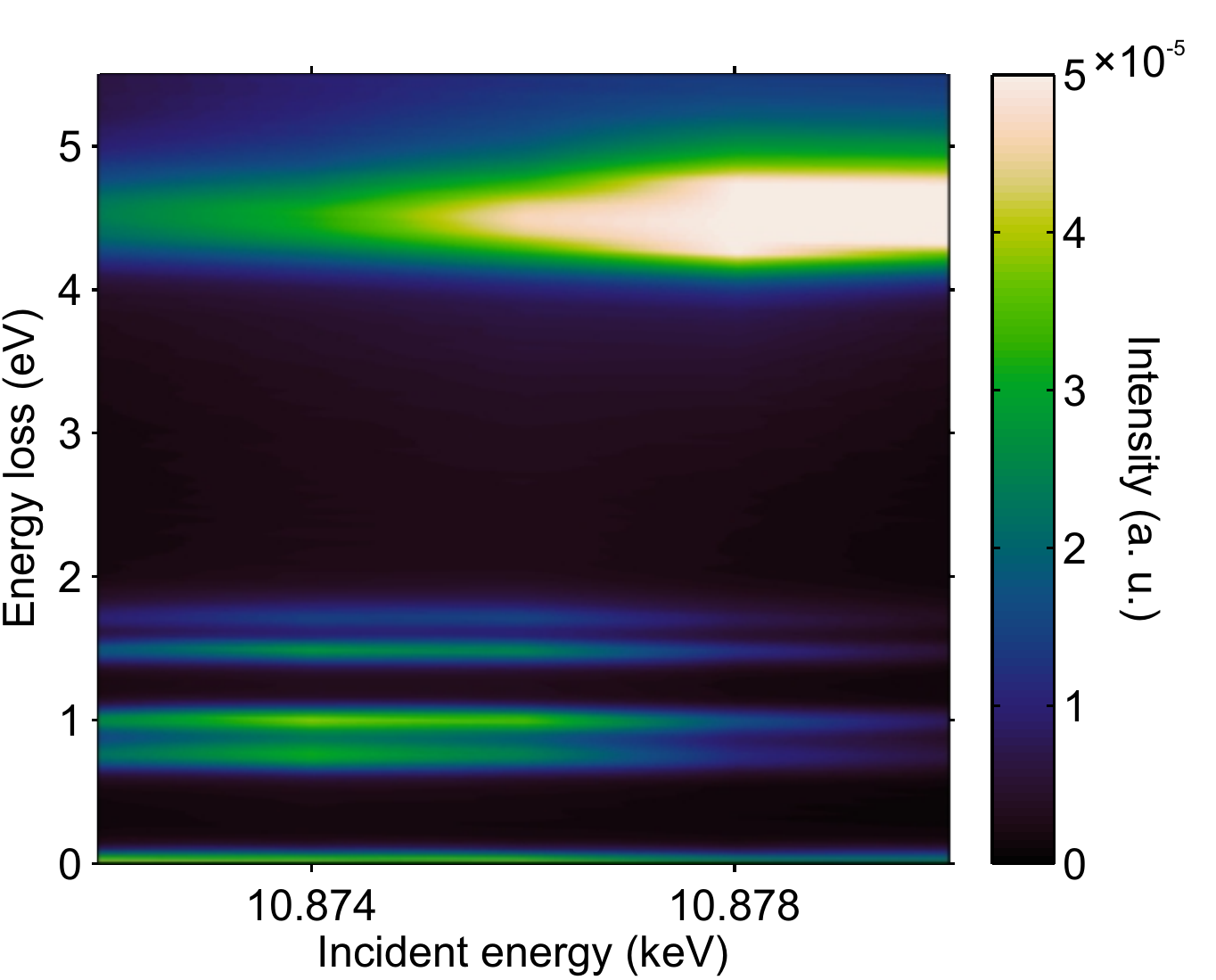}
\protect\caption{\label{fig:colormap}Incident energy dependence of electronic excitations
in \CLOO{}. Measurements were performed at 300\,K.\hspace{1cm}}
\end{figure}

Figure~\ref{fig:colormap} presents the x-ray energy loss, $E$,
versus incident energy, $E_{\mathrm{i}}$, RIXS spectra of \CLOO{}
at 300$\,$K. Four lines are present at $E<2\,$eV, which are
enhanced at $E_{\mathrm{i}}=10.874\,$keV, whereas the feature at
$E\approx4.5\,$eV is enhanced at $E_{\mathrm{i}}=10.878\,$keV. This
indicates that the $E<2\,$eV features are intra-$t_{2g}$ excitations,
whereas the higher energy feature is from $t_{2g}$ to $e_{g}$ excited
states, as has been observed in many $5d$ oxides~\cite{liu_testing_2012,boseggia_antiferromagnetic_2012,sala_cairo3:_2014,calder_spin-orbit-driven_2016}.
Subsequent measurements were optimised to probe the $t_{2g}$ excitations
by fixing $E_{\mathrm{i}}=10.874\,$keV. 

\begin{figure}[t!]
\includegraphics[width=0.9\columnwidth]{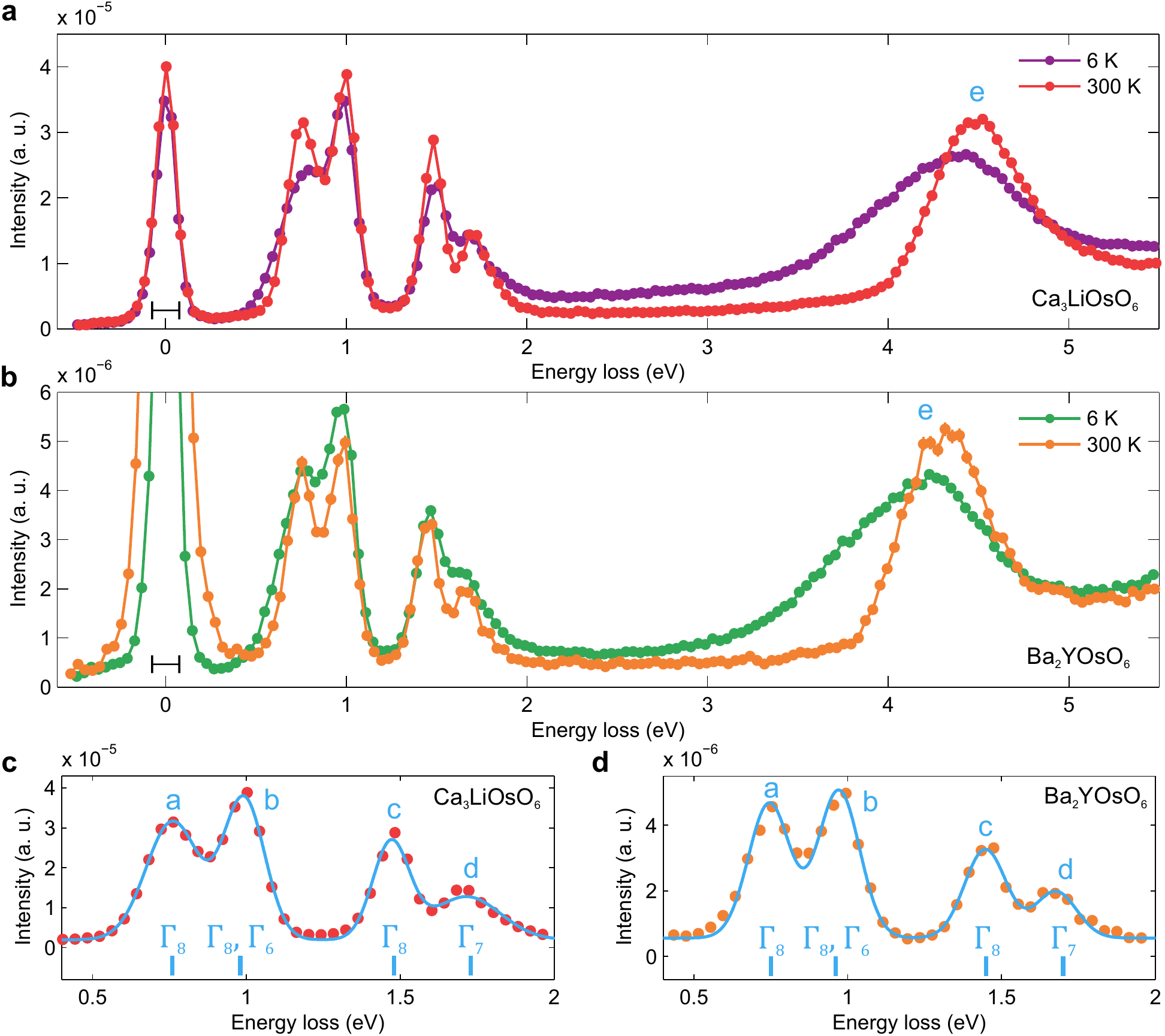}
\protect\caption{\label{fig:RIXS-excitation-spectra}RIXS excitation spectra measured
in \CLOO{} and \BYOO{}. Panels a and b show the data for \CLOO{}
and \BYOO{}, respectively. Black lines indicate the full-width half
maximum of the elastic line. Panels c and d show the energy range
0.4--2\,eV, i.e. the $t_{2g}$ manifold, with the results of Gaussian
peak fitting to the data shown as solid lines. }
\end{figure}

Figure~\ref{fig:RIXS-excitation-spectra} presents the detailed RIXS
spectra of \CLOO{} and \BYOO{} at temperatures of 300 K and 6 K.
In each spectrum there are 5 peaks in addition to the elastic line:
four peaks with $E<2\,$eV, labeled a, b, c and d (Fig.~\ref{fig:RIXS-excitation-spectra}c
and~d) which we associated with intra-$t_{2g}$ excitations, and
a broad peak, e, centered at $E\sim4.5\,$eV (Fig.~\ref{fig:RIXS-excitation-spectra}a
and~b) associated with $t_{2g}$ to $e_{g}$ excited states. The
qualitative similarity of the spectra indicates that these features
are not controlled by non-cubic structural distortions, as splittings would be heightened in \CLOO{}.

At 300\,K the peaks a--d are resolution limited, as determined
from least-squares fitting of Gaussian peaks on a flat background
to the data, Fig.~\ref{fig:RIXS-excitation-spectra}c
and~d. The peak energies for \CLOO{} are $a_{\mathrm{Ca}}=0.760(7)\,$eV,
$b_{\mathrm{Ca}}=0.992(5)\,$eV, $c_{\mathrm{Ca}}=1.470(5)\,$eV and
$d_{\mathrm{Ca}}=1.72(1)\,$eV, and for \BYOO{} are $a_{\mathrm{Ba}}=0.745(7)\,$eV,
$b_{\mathrm{Ba}}=0.971(7)\,$eV, $c_{\mathrm{Ba}}=1.447(9)\,$eV and
$d_{\mathrm{Ba}}=1.68(1)\,$eV. At 6$\,$K, well below $T_\mathrm{N}$ in both materials, the peaks appear broadened,
although maintain the SOC-induced four peak character, see Fig.~\ref{fig:RIXS-excitation-spectra}.
The low temperature broadening is most pronounced in peak e --- this
peak is due to 20 different excited levels in the Coulomb plus
SOC regime, so no discrete levels can be resolved with current RIXS
capabilities. The broadening could be
due to splittings from non-cubic structural distortions occurring
below the magnetic ordering temperatures - some distortion should
be expected from magnetoelastic coupling. However, if a purely structural
distortion were the origin we would expect to see broadening in \CLOO{}
compared to \BYOO{} at all temperatures. It is possible that at low
temperatures there is dispersion of the levels~\cite{ament_resonant_2011},
or that increased hybridisation influences the accessible distribution
of excited states --- this would be more pronounced for $e_{g}$
levels which show greater oxygen overlap. Here, we focus on what the
splitting of the $t_{2g}$ character peaks reveals about the electronic
ground state.

\begin{figure}[t!]
\includegraphics[width=0.9\columnwidth]{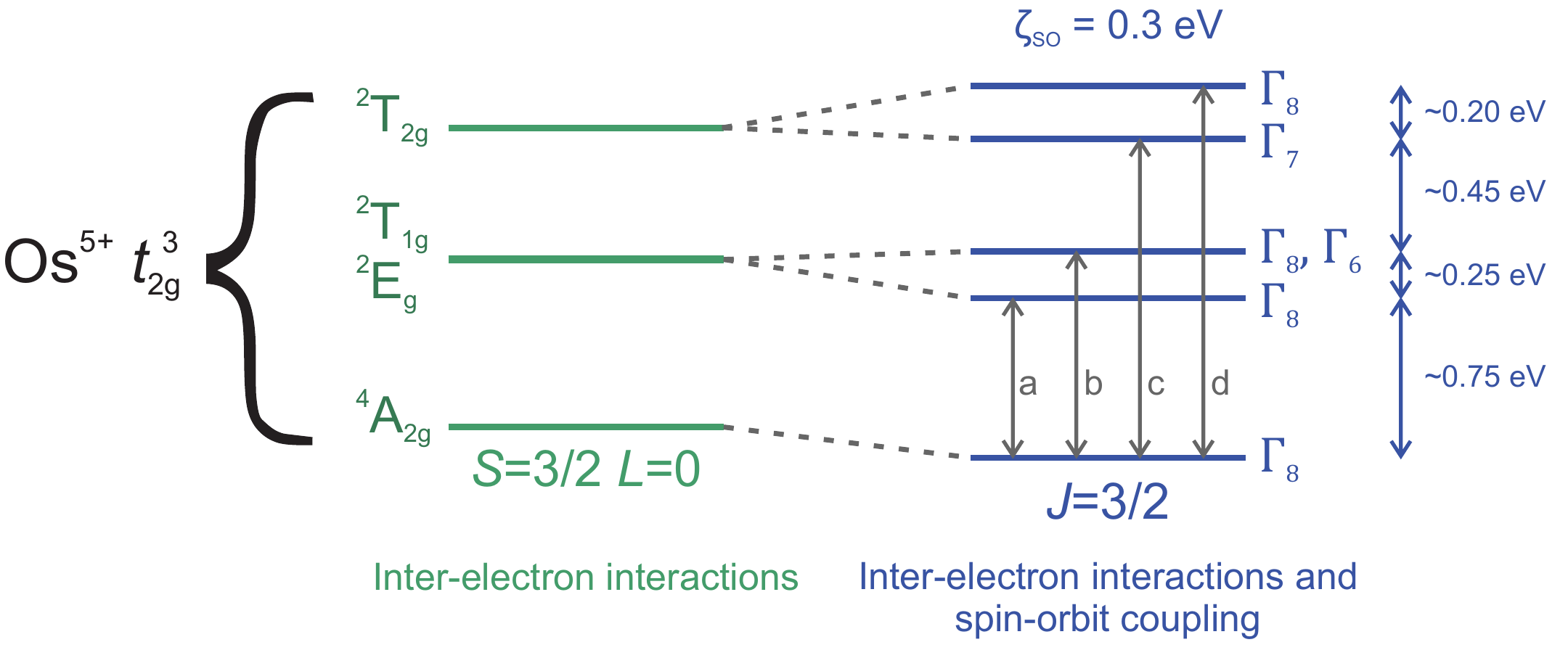}
\protect\caption{\label{fig:levels}Structure of $t_{2g}^{3}$ levels for Os$^{5+}$
ions in octahedral environment with strong spin-orbit interaction.
The irreducible representations for the states without SOC, i.e. $\zeta_{\mathrm{SO}}=0$,
are labeled by the appropriate Mulliken Symbols, full spatial forms
are tabulated in many textbooks e.g. Ref.~\cite{sugano_multiplets_1970}. Here $^2E_g$ and $^2T_{2g}$ appear degenerate as we determine $B=0.00(4)\,$eV within resolution, see main text.
The irreducible representations describing the SOC-induced states
are four-fold degenerate, $\Gamma_{8}$, and two-fold degenerate,
$\Gamma_{6}$ and $\Gamma_{7}$. Dashed lines link the final states
with the $\zeta_{\mathrm{SO}}=0$ states which provide the greatest
contribution to them, however each final state is an intermixing of
all $\zeta_{\mathrm{SO}}=0$ states, which allows SOC to enter the
ground state $\Gamma_{8}$. The labels a, b, c and d indicate the
excited-state energies that are observed in the RIXS spectra of \BYOO{}
and \CLOO{}. }
\end{figure}

 The four $t_{2g}$-character peaks we observe, Fig.~\ref{fig:RIXS-excitation-spectra}, are incompatible with the multiplets expected in standard crystal field theory developed for 3$d$ ions which leads to the S=3/2 ground state~\cite{sugano_multiplets_1970}. Here, we identify that a method first proposed by Kamimura \emph{et
al.}~\cite{kamimura_magnetic_1960} for transition metal halides is relevant in this case, in which the assumption that inter-electron interaction energies (including
intra- and inter-orbital Coulomb and exchange interactions) are much
larger than SOC is dropped (a similar method was also recently derived in Ref. \cite{matsuura_effect_2013}).  We identify that such an approach can be utilised to model our TMO RIXS data, with the primary assumption that the hybridisation with the surrounding
oxygen ligands leaves the transformation properties of the free ion
wavefunctions unaltered. This therefore promotes the breaking of the $S=3/2$ singlet and strong entry of SOC. The wavefunctions
are therefore described in terms of irreducible representations in
the $O$ double group determined by the octahedral symmetry~\cite{eisenstein_magnetic_1961,brian_n._figgis_ligand_2000,sugano_multiplets_1970},
Fig.~\ref{fig:levels}. This formulation does not necessitate approximations that
the cubic crystal field is infinite or that Coulomb or SOC must be neglected, and is not restricted to use for $5d^{3}$ configurations~\cite{kamimura_magnetic_1960}.

To determine the wavefunctions for the 5$d^{3}$ case, as in crystal field theory we use initial basis states that describe the ways in which three electrons can occupy the pure $t_{2g}$ and $e_{g}$ levels (i.e. terms such as $^{4}$A$_{2\mathrm{g}}$ - which
is one electron in each of $d_{xy}$, $d_{yz}$, $d_{xz}$ with total
spin 3/2), and then apply inter-electron
interaction and SOC between these states. The inter-electron interactions are expressed in terms of the Racah parameters $B$ and $C$. Due to hybridisation, the radial form of the $d$ levels are unknown, i.e. $B$ and $C$ deviate from pure ionic values, but the Racah parameters are formulated such that they can be easily extracted from experiment~\cite{brian_n._figgis_ligand_2000,sugano_multiplets_1970,georges_strong_2013}.
The same interactions can be expressed in terms of intra- and inter-orbital
Coulomb interactions, $U$ and $U^{\prime}$, and the effective Hund's
coupling $J_{h}$, which have the relationships to the Racah parameters
$J_{h}=3B+C$, $U=A+4B+3C$ and $U^{\prime}=A-2B+C$~\cite{georges_strong_2013}.
The Racah parameter $A$, however, only appears in the diagonal matrix
elements of the interactions matrices and causes a constant shift
in all eigenvalues, including the ground state,~\cite{eisenstein_magnetic_1961,dorain_optical_1966}
so cannot be determined by the energies of the excited states probed
by RIXS. The full interaction matrices are given in
Ref.~\cite{eisenstein_magnetic_1961}. By numerically diagonalising
the matrices we determine the eigenstates illustrated in Fig.~\ref{fig:levels},
and are able to fit the resulting eigenvalues to the determined energies $a_{\mathrm{Ba}}$--$d_{\mathrm{Ba}}$ and $a_{\mathrm{Ca}}$--$d_{\mathrm{Ca}}$,
Fig.~\ref{fig:RIXS-excitation-spectra}, see supplementary material
for further details. We fix the value of the crystal field to the
peak value of peak e, $10Dq=4.5\,$eV for \CLOO{} and $10Dq=4.3\,$eV
for \BYOO{}, because the positions of levels $a$--$d$ are insensitive
to small changes in this term, and the resulting levels include negligible
mixing of $e_{g}$ states, as expected for a strong cubic crystal
field splitting.  

The resulting parameters provide direct insight into the dominant
interactions in the materials. For \CLOO{} we find $\zeta_{\mathrm{SOC}}=0.35(7)\,$eV,
$B=0.00(5)$, $C=0.3(2)$ and  $J_{h}=3B+C=0.3(2)\,$eV, and for \BYOO{}
we find $\zeta_{\mathrm{SO}}=0.32(6)\,$eV, $B=0.00(5)\,$eV, $C=0.3(2)\,$eV
and $J_{h}=0.3(2)\,$eV. The fact that $\zeta_{\mathrm{SO}}$
is of similar size to $C$ (and $J_{h}$) clearly demonstrates that
perturbative approaches are not appropriate for the treatment of SOC
in $5d^{3}$ systems. The ratio of $C$/$B$ is commonly used to indicate
the scale of deviation from pure ionic wavefunctions (the nephelauxetic effect) by comparison to the same
ratio in 3$d$ materials where SOC is weak, with $C/B=4$ for Cr$^{3+}$
(3$d^{3}$) ions~\cite{brian_n._figgis_ligand_2000,sugano_multiplets_1970}.
We find $B$ to be zero within the error, indicative of a
large nephelauxetic effect;
improved resolution of RIXS measurements would be advantageous for
this comparison. We can, however, look at the eigenvector we determined
for the $\Gamma_{8}$ ground state - which is a linear combination
of the 21 initial basis states which describe 3 electrons occupying
the $t_{2g}$ levels, or two occupying the $t_{2g}$ levels and one
in the $e_{g}$ levels (the latter ultimately form a negligible contribution,
see supplementary material). 
The largest component is, as expected,
from the $S=3/2$ $\left|^{4}A_{2}\right\rangle $ state, but the
next major contribution is from the $S=1/2$ $\left|^{2}T_{2g}\right\rangle $
state. 
For \CLOO{} (\BYOO{}), these  $\left|^{4}A_{2}\right\rangle $ and $\left|^{2}T_{2g}\right\rangle $ terms appear in the normalised eigenvector with weights of 0.95 (0.95) and 0.27 (0.25), respectively --- the complete eigenfunctions are given in the supplementary material. This latter component (plus smaller terms) directly explains the entry
of SOC physics and the observations of small orbital moments for $5d^{3}$
ions~\cite{taylor_magnetic_2015,taylor_spin-orbit_2016,veiga_fragility_2015,laguna-marco_electronic_2015,kermarrec_frustrated_2015}.

We finally explore how the framework presented provides insight about the physical manifestation
of SOC. An x-ray absorption near-edge spectroscopy
plus x-ray circular dichroism study of 5$d^{3}$ Ir$^{6+}$ double
perovskites found strong coupling between orbital and spin moments
despite small orbital moments, and suggested this should be due to
some deviation from the pure $t_{2g}^3$ levels~\cite{laguna-marco_electronic_2015},
with similar results reported in Os$^{5+}$ materials~\cite{veiga_fragility_2015}.
The wavefunction we determine explains these results, with a $J=3/2$
state which has only a small orbital moment. The observation of a
large spin gap in the magnetic excitation spectra of \BYOO{} and
related double perovskites is also explained by the intermediate coupling
framework, as the spin gap results from strong SOC-induced anisotropy
which is unexplained in a $S=3/2$ picture. In these double perovskites
and the pyrochlore Cd$_{2}$Os$_{2}$O$_{7}$ anisotropy is held responsible
for stabilisation of the observed magnetic ground states~\cite{kuzmin_effect_2003,taylor_spin-orbit_2016,calder_spin-orbit-driven_2016}.
Our results therefore show that intermediate coupling electronic ground states are central in dictating the macroscopic physical properties of such materials. We speculate that a similar approach should be utilised for 5$d^{2}$ and 5$d^{4}$ materials, as they are expected to show larger SOC effects~\cite{chen_spin-orbit_2011}, alongside increased hybridisation in the $d^{2}$ case. The $d^{4}$ materials have attracted interest for hosting magnetic order~\cite{cao_novel_2014}
despite the $d{}^{4}$ configuration leading to non-magnetic singlets
in either LS or jj limits~\cite{chen_spin-orbit_2011}.

\noindent
\\
\textbf{Acknowledgements}

We thank A. Huq and M. J. Kirkham for assistance with high-resolution
neutron and x-ray diffraction experiments. We thank G. Pokharel for useful discussions. Use of the Advanced Photon
Source at Argonne National Laboratory was supported by the U. S. Department
of Energy, Office of Science, Office of Basic Energy Sciences, under
Contract No. DE-AC02-06CH11357. The research at ORNL's Spallation
Neutron Source and High Flux Isotope Reactor was supported by the
Scientific User Facilities Division, Office of Basic Energy Sciences,
U.S. Department of Energy (DOE). This research was supported in part
by the Center for Emergent Materials an National Science Foundation
(NSF) Materials Research Science and Engineering Center (DMR-1420451).
This research was supported in part by the Japan Society for the Promotion
of Science (JSPS) through a Grant-in-Aid for Scientific Research (15K14133,
16H04501). The authors declare no competing financial interests. This manuscript has been authored by UT-Battelle, LLC under Contract No. DE-AC05-00OR22725 with the U.S. Department of Energy.  The United States Government retains and the publisher, by accepting the article for publication, acknowledges that the United States Government retains a non-exclusive, paid-up, irrevocable, world-wide license to publish or reproduce the published form of this manuscript, or allow others to do so, for United States Government purposes.  The Department of Energy will provide public access to these results of federally sponsored research in accordance with the DOE Public Access Plan (http://energy.gov/downloads/doe-public-access-plan).


\clearpage

\begin{center}
{\LARGE Supplementary Material for: \\}
{\LARGE Spin-orbit coupling controlled $J=3/2$
electronic ground state in 5$d^{3}$ oxides}
\end{center}

\noindent
\\
\textbf{Resonant inelastic x-ray scattering} 

RIXS measurements were
performed at the Advanced Photon Source (APS) at Sector 27 using the
MERIX instrumentation~\cite{gog_momentum-resolved_2009}. A closed-cycle
refrigerator was used to control the sample temperature. The Os L$_{3}$-edge
incident energy was accessed with a primary diamond(1 1 1) monochromator
and a secondary Si(4 0 0) monochromator. A Si(4 6 6) diced analyzer
was used to determine the energy of the beam scattered from the sample,
and a MYTHEN strip detector was utilised. All measurements were performed
with $2\mathrm{\theta}=90^{\circ}$ in horizontal geometry. The RIXS
energy resolution was 150$\,$meV FWHM. The raw data counts are normalised
to the incident beam intensity via an ion chamber monitor. To compare
the temperatures, the 6\,K data is normalised so that the featureless
energy gain side overlaps with the 300\,K data \textendash{} this
accounts for deviations of the beam position on the sample. 

\noindent
\\
\textbf{Synthesis} 

The synthesis route and properties of \CLOO{}
are reported in Ref.~\cite{shi_crystal_2010}. Characterisation of
the powder sample utilised here, including diffraction measurements
and magnetic structure determination, are reported in Ref.~\cite{calder_magnetic_2012}.
The powder sample of \BYOO{} of mass 1.5 g was synthesised by grinding
together stoichiometric quantities of barium peroxide, yttrium oxide,
and osmium metal. The reactants were loaded into an alumina tube which
was placed into a silica tube along with a secondary alumina vessel
containing PbO$_{2}$. The silica tube was sealed under dynamic vacuum,
and heated to 1000$\,^{\circ}$C for a period of 48\,hours in a box
furnace located in a fume hood. The PbO$_{2}$ decomposed into PbO
at elevated temperature, acting as a source of O$_{2}$ gas for the
reaction. A calculated excess was used resulting in $\frac{1}{4}$
mole excess O$_{2}$ per mole of product in order to ensure full oxidation
of the reactants. Care must be taken when heating Os or Os containing
compounds due to the potential formation of highly toxic OsO$_{4}$
gas. X-ray and neutron diffraction and susceptibility measurements
were consistent with those previously reported~\cite{kermarrec_frustrated_2015},
and are detailed in the supplementary material.

\noindent
\\
\textbf{Ca$_{3}$LiOsO$_{6}$ Characterisation}

Polycrystalline \CLOO{} was used in this work, for which synthesis
route and properties are reported in Ref.~\cite{shi_crystal_2010}. Neutron powder
diffraction measurements and magnetic structure determination are
reported in Ref.~\cite{calder_magnetic_2012}.

\noindent
\\
\textbf{Ba$_{2}$YOsO$_{6}$ Characterisation}

\begin{figure}
\includegraphics[width=0.5\columnwidth]{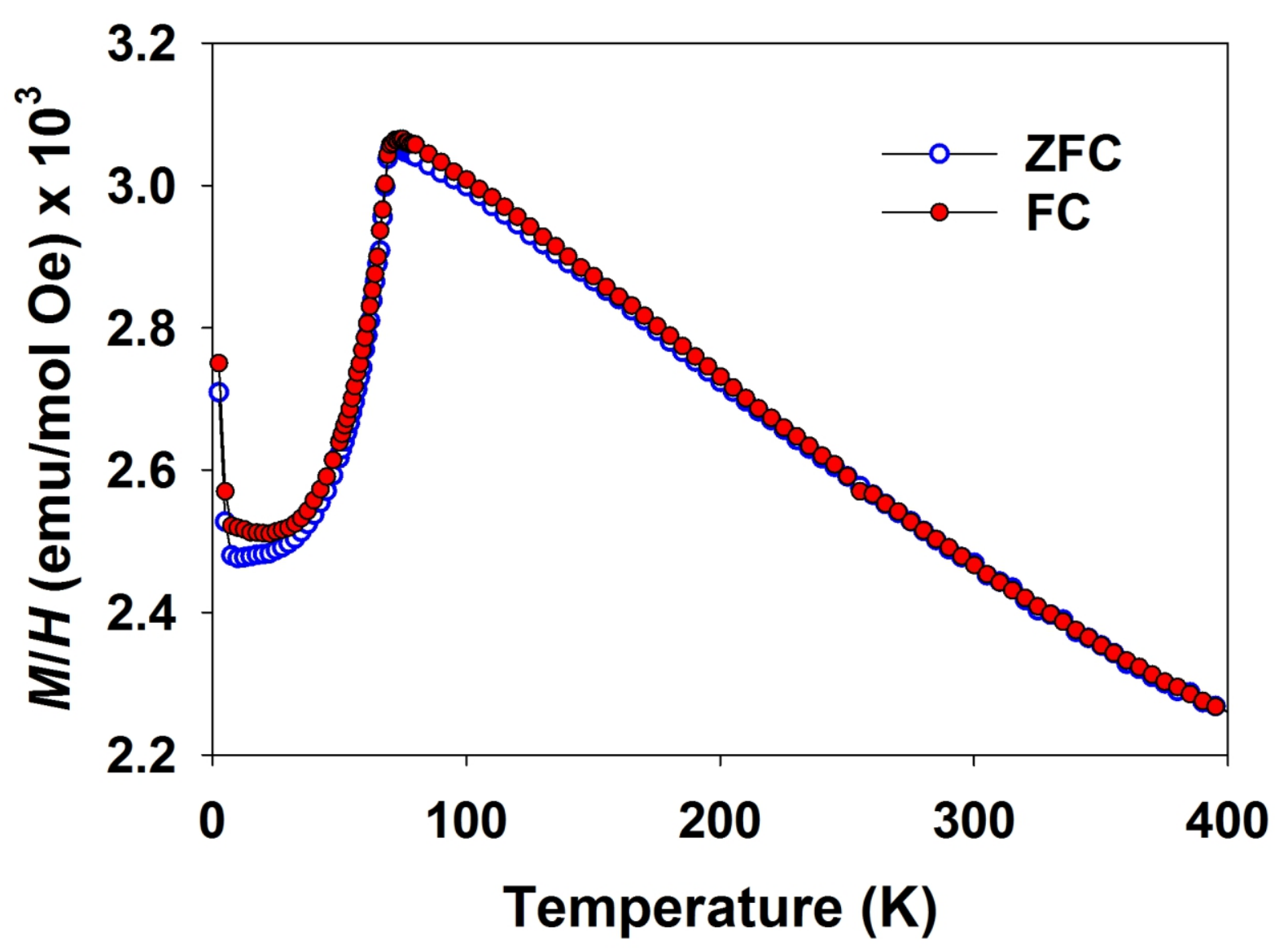}\protect\caption{\label{fig:squid}Temperature dependence of the zero field cooled
(blue open circles) and field cooled (red closed circles) magnetization
of \BYOO{} under 1\,kOe measurement field. The low temperature rise
of the magnetization is attributed to the presence of ferromagnetic
impurity  Ba$_{11}$Os$_4$O$_{24}$, which has $T_{\mathrm{C}}=6.8\,$K\cite{wakeshima_crystal_2005}.}
\end{figure}

The temperature dependence of the magnetization of the \BYOO{} powder
was measured with a Quantum Design MPMS SQUID magnetometer. The powder
was contained in a gel capsule and mounted in a straw for insertion
into the device. Data were collected between 2.5 and 400\,K under
an applied field of 1\,kOe under field cooled (FC) and zero-field
cooled (ZFC) conditions. No corrections were applied to the data.
A sharp transition at 70\,K was observed in both data sets, see Fig.\ref{fig:squid},
associated with the antiferromagnetic ordering of Os$^{5+}$ as previously
observed\cite{kermarrec_frustrated_2015}. The rise of the magnetization
at very low temperatures is attributed to the presence of ferromagnetic
impurity  Ba$_{11}$Os$_4$O$_{24}$, which has $T_{\mathrm{C}}=6.8\,$K\cite{wakeshima_crystal_2005}.
A Curie-Weiss fit was conducted in the temperature range 300--400\,K
resulting in a Weiss constant, $\Theta$, of -785\,K and an effective
moment of 4.27$\,\upmu\mathrm{B}$. These parameters result in a large
frustration index, $|\Theta/T_{\mathrm{N}}|$, of 11.2. However, the
effective moment is larger than the predicted spin-only moment of
3.87$\,\upmu_{B}$ suggesting that the true paramagnetic regime is
not achieved below 400\,K, consistent with Ref.~\cite{kermarrec_frustrated_2015}.

\noindent
\\
\textbf{Ba$_{2}$YOsO$_{6}$ Structural Study}

High resolution x-ray and neutron diffraction experiments were performed
on \BYOO{} in an attempt to identify any non-cubic distortion. The
synchrotron x-ray experiments are optimised for identifying non-cubic
lattice parameter splittings, and for identifying Y-Os site mixing,
whereas the neutron scattering experiments are more likely to be able
to identify distortions and tilting of the oxygen octahedra. Space
group \emph{Fm}$\bar{3}$\emph{m} was used for the refinements, in
which the ions are on Wyckoff sites: Ba 8\emph{c} $(\frac{1}{4},\frac{1}{4},\frac{1}{4})$,
Y 4\emph{a} $(0,0,0)$, Os 4\emph{b} $(\frac{1}{2},\frac{1}{2},\frac{1}{2})$
and O1 24\emph{e}, $(0, 0, z)$. The Y and Os ions contribute to the
same Bragg reflections, therefore for both x-ray and neutron refinements
the displacement parameters for these ions were constrained to be
equal.

\begin{figure}
\includegraphics[width=0.98\columnwidth]{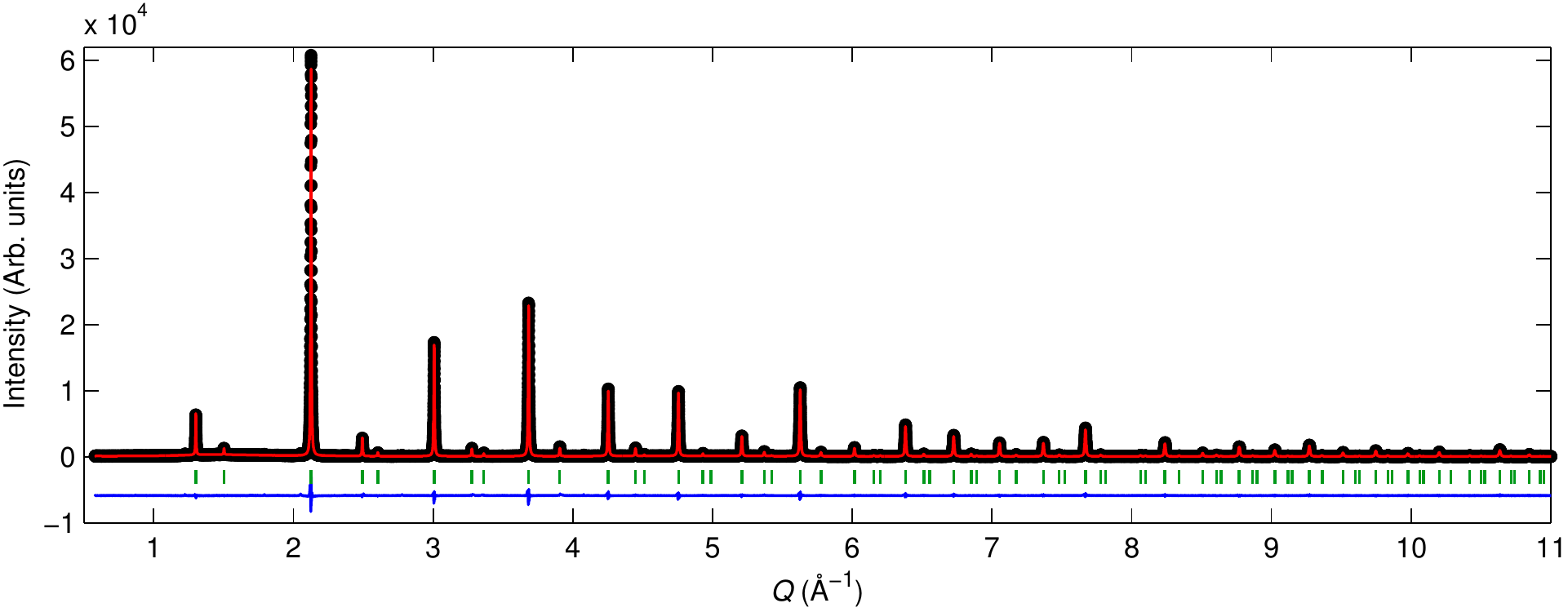}\protect\caption{\label{fig:xrpd}X-ray powder diffraction pattern of \BYOO{} measured
at 295\,K. The black circles are observed data, the red line is the
calculated pattern, and the blue line is the difference of the two.
Allowed reflections of the phase are given as green hashes. }
\end{figure}

Synchrotron x-ray diffraction measurements were conducted on 11-BM
at the Advanced Photon Source with wavelength $\lambda=0.4593\,\mathrm{\AA}$
at temperature of 295\,K. The 0.012$\,$g of sample was mixed with
ground quartz in a 1:1 mass ratio to minimize absorption, and this
was placed in a 0.4mm radius capillary for the measurement. The data
were analyzed via Rietveld refinement as implemented in GSAS\cite{toby_expgui_2001}.
Y/Os cation ordering refinements were performed indicating full cation
ordering, similar to previous work\cite{kermarrec_frustrated_2015}.
No non-cubic peak splittings, nor additional peaks, could be identified.
The results of refinements based on space group $Fm\bar{3}m$ is illustrated
in Fig.\ref{fig:xrpd} and summarized in Table~\ref{tab:xrpd}. 

\begin{table}
\protect\caption{\label{tab:xrpd}Results from Rietveld refinements against XRPD data
from BM-11 using GSAS. Errors quoted in parenthesis are the standard
deviations determined by GSAS. }
\begin{tabular}{|c|c|c|c|c|c|c|c|}
\hline 
Temperature & $a(\mathrm{\AA})$ & $\chi^{2}$ & $R_{wp}$ & O1 $z$ & Ba $B_{iso}$ & Y/Os $B_{iso}$ & O1 $B_{iso}$\tabularnewline
\hline 
\hline 
295\,K & 8.35518(1) & 2.534  & 8.75$\,$\% & 0.2379(2) & 0.709(4) & 0.248(3) & 1.00(3)\tabularnewline
\hline 
\end{tabular}
\end{table}

\begin{figure}
\includegraphics[width=0.98\columnwidth]{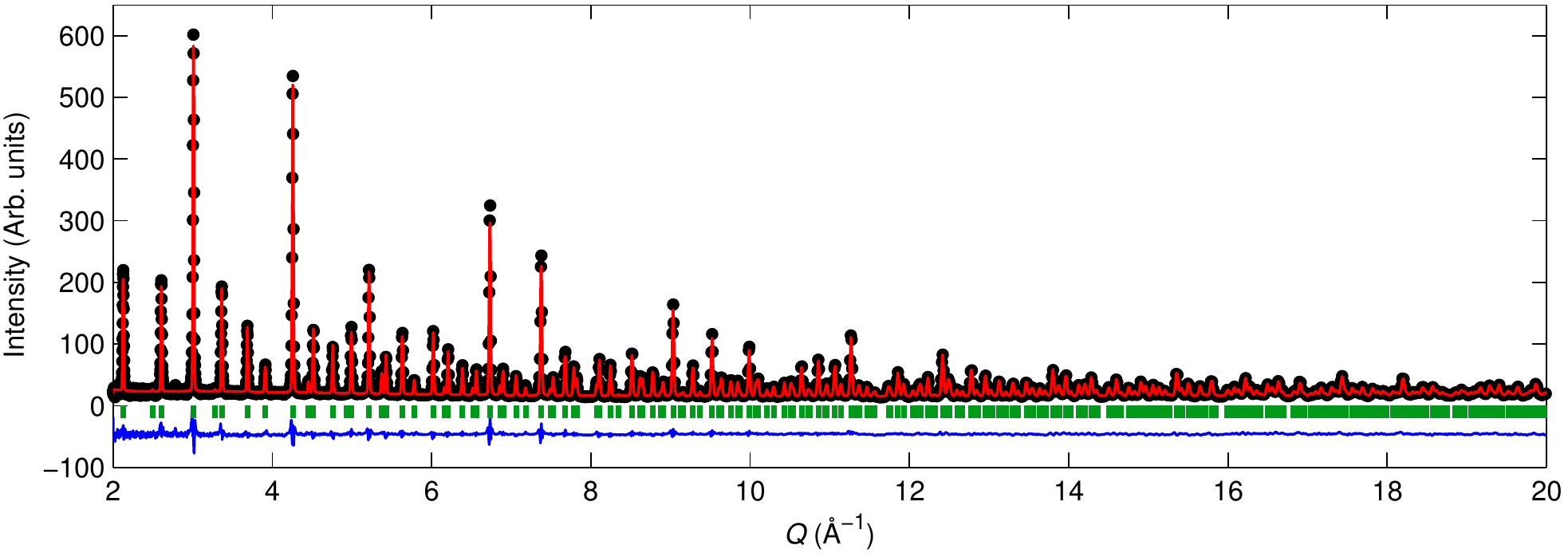}

\protect\caption{\label{fig:npd}Neutron powder diffraction pattern of \BYOO{} measured
at 10\,K. The black circles are observed data, the red line is the
calculated pattern, and the blue line is the difference of the two.
Allowed reflections of the phase are given as green hashes. }
\end{figure}

Neutron powder diffraction measurements were conducted on POWGEN at
the Spallation Neutron Source at Oak Ridge National Laboratory (ORNL).
1.4\,g of \BYOO{} was contained in a vanadium can, and measured
at 10 and 100\,K. High-resolution settings were chosen, $\lambda=1.066\,\mathrm{\AA}$
and $f=60\,$Hz, to optimize chances of identifying a non-cubic distortion.
The data were analyzed via Rietveld refinement as implemented in Fullprof\cite{rodriguez-carvajal_recent_1993}.
No peaks associated with octahedral rotations in non-cubic symmetries
could be identified\cite{howard_ordered_2003}. The results of refinements
based on space group $Fm\bar{3}m$ is illustrated in Fig.\ref{fig:npd}
and summarized in Table~\ref{tab:npd}.
In this refinement all ion occupancies were kept at 100\,\% as indicated
by the x-ray data. Using anisotropic displacement parameters for the
oxygen positions did not produce an improvement to the fit qualities,
so we quote the results with isotropic displacement parameter results
for all ions. 

\begin{table}
\begin{tabular}{|c|c|c|c|c|c|c|c|c|}
\hline 
Temperature & $a(\mathrm{\AA})$ & $\chi^{2}$ & $\mathrm{R_{Bragg}}$ & $\mathrm{R_{F}}$ & O1 $z$ & Ba $B_{iso}$ & Y/Os $B_{iso}$ & O1 $B_{iso}$\tabularnewline
\hline 
\hline 
10\,K & 8.34383(8) & 9.91 & 6.76 & 3.77 & 0.23446(5) & 0.104(7) & 0.107(5) & 0.293(6)\tabularnewline
\hline 
100\,K & 8.34596(8) & 10.1 & 6.83 & 4.10 & 0.23438(5)  & 0.183(7) & 0.132(6) & 0.342(7)\tabularnewline
\hline 
\end{tabular}

\protect\caption{\label{tab:npd}Results from Rietveld refinements against NPD data
from POWGEN using Fullprof. Errors quoted in parenthesis are the standard
deviations determined by Fullprof. }

\end{table}

\noindent
\\
\textbf{Spin-Orbit Calculation}

As referenced in the main text, the complete Hamiltonian describing
cubic-crystal field, Coulomb interactions (including Hund's coupling)
and spin-orbit effects is given by Eisenstein\cite{eisenstein_magnetic_1961}
in the form of a 21$\times$21 matrix for $\Gamma_{8}$ and a $9\times9$
matrix for each of the $\Gamma_{6}$ and $\Gamma_{7}$ representations,
describing the interactions between each of the basis states utilized
in the standard Coulomb-only model\cite{kanamori_electron_1963,sugano_multiplets_1970,georges_strong_2013}.
These basis states are the $\left|^{4}A_{2}\right\rangle $ , $\left|^{2}T_{2g}\right\rangle $
etc. states which describe all the ways in which three electrons can
occupy the $t_{2g}$ and $e_{g}$ levels, and are documented in full
in textbooks, for example Ref.~\cite{sugano_multiplets_1970}. The
interaction matrices are formulated in terms of the Racah parameters,
$A_{n}$, $B_{n}$ and $C_{n}$ ($n=0$ to $4$ is symmetry allowed)
but we follow the formulation in Ref.~\cite{dorain_optical_1966}
and assume only one parameter of each type is required (i.e. $A_{n}=A$,
$B_{n}=B$, $C_{n}=C$ $\forall n$) as is commonly adopted\cite{georges_strong_2013},
and found to be reasonable in Ref.~\cite{eisenstein_magnetic_1961}
for Re$^{4+}$ in Cl octahedra. The term $A$ is only found on the
diagonal elements, adding 3$A$ to each eigenvalue, and therefore
the differences in energy between states is independent of $A$ and
we can set it to any arbitrary value in describing our data. We diagonalise
the matrices to find the eigenvalues and eigenvectors, and shift all
eigenvalues by the energy of the lowest term, so that the ground state
is at $E=0$. We then fit the first four excited state eigenvalues
to the observed RIXS excitations as described in the main text. 

The basis vectors of the $\Gamma_{8}$ representation in the O double
group are angular momentum wavevectors $\phi(J,m_{J})$ with $J=3/2$
and $m_{J}=\frac{3}{2}$, $\frac{1}{2}$, $-\frac{1}{2}$ and $-\frac{3}{2}$\cite{dresselhaus_group_2008,koster_properties_1963}.
We determine the eigenvector for the $\Gamma_{8}$ ground state in
terms of a linear combination of 21 basis states describing the three
electron occupations of the $t_{2g}$ and $e_{g}$ levels, including
$t_{2g}$-$e_{g}$ excited states, although as $10Dq\sim4\,$eV, the
contributions from these states are extremely small. 

The complete normalized eigenfunction of the ground state for \BYOO{}
from a numerical diagonalisation is, to 3 decimal places: \begin{equation}
\begin{split}
\left|\Gamma_{8}\mathrm{g.s.}\right\rangle ={} 
&-0.953\left|^{4}A_{2}(t_{2g}{}^{3})\right\rangle +0.069\left|^{2}E_{g}(t_{2g}{}^{3})\right\rangle +0.017\left|^{2}E_{g}(t_{2g}{}^{2}e_{g}{}^{1})\right\rangle +0.015\left|^{2}E_{g}(t_{2g}{}^{2}e_{g}{}^{1})\right\rangle 
\\&-0.001\left|^{2}E_{g}(e{}^{3})\right\rangle -0.081\left|^{2}T_{1g}(t_{2g}{}^{3})\right\rangle -0.007\left|^{2}T_{1g}(t_{2g}{}^{2}e_{g}{}^{1})\right\rangle +0.013\left|^{2}T_{1g}(t_{2g}{}^{2}e_{g}{}^{1})\right\rangle 
\\&-0.004\left|^{2}T_{1g}(t_{2g}{}^{1}e_{g}{}^{2})\right\rangle +0.002\left|^{2}T_{1g}(t_{2g}{}^{1}e_{g}{}^{2})\right\rangle -0.024\left|^{4}T_{1g}(t_{2g}{}^{2}e_{g}{}^{1})\right\rangle
\\&+0.006\left|^{4}T_{1g}(t_{2g}{}^{1}e_{g}{}^{2})\right\rangle -0.007\left|^{4}T_{1g}(t_{2g}{}^{2}e_{g}{}^{1})\right\rangle -0.002\left|^{4}T_{1g}(t_{2g}{}^{1}e_{g}{}^{2})\right\rangle \\&-0.252\left|^{2}T_{2g}(t_{2g}{}^{3})\right\rangle  +0.069\left|^{2}T_{2g}(t_{2g}{}^{2}e_{g}{}^{1})\right\rangle -0.013\left|^{2}T_{2g}(t_{2g}{}^{2}e_{g}{}^{1})\right\rangle \\&-0.011\left|^{2}T_{2g}(t_{2g}{}^{1}e_{g}{}^{2})\right\rangle    +0.006\left|^{2}T_{2g}(t_{2g}{}^{1}e_{g}{}^{2})\right\rangle -0.038\left|^{4}T_{2g}(t_{2g}{}^{2}e_{g}{}^{1})\right\rangle \\&-0.098\left|^{4}T_{2g}(t_{2g}{}^{2}e_{g}{}^{1})\right\rangle .
\end{split}
\end{equation} The complete normalized eigenfunction of the ground state for \CLOO{}
from a numerical diagonalisation is, to 3 decimal places: \begin{equation}
\begin{split}
\left|\Gamma_{8}\mathrm{g.s.}\right\rangle ={} 
&-0.947\left|^{4}A_{2}(t_{2g}{}^{3})\right\rangle +0.076\left|^{2}E_{g}(t_{2g}{}^{3})\right\rangle +0.019\left|^{2}E_{g}(t_{2g}{}^{2}e_{g}{}^{1})\right\rangle +0.016\left|^{2}E_{g}(t_{2g}{}^{2}e_{g}{}^{1})\right\rangle 
\\&-0.001\left|^{2}E_{g}(e{}^{3})\right\rangle -0.090\left|^{2}T_{1g}(t_{2g}{}^{3})\right\rangle -0.007\left|^{2}T_{1g}(t_{2g}{}^{2}e_{g}{}^{1})\right\rangle +0.014\left|^{2}T_{1g}(t_{2g}{}^{2}e_{g}{}^{1})\right\rangle 
\\&-0.004\left|^{2}T_{1g}(t_{2g}{}^{1}e_{g}{}^{2})\right\rangle +0.002\left|^{2}T_{1g}(t_{2g}{}^{1}e_{g}{}^{2})\right\rangle -0.025\left|^{4}T_{1g}(t_{2g}{}^{2}e_{g}{}^{1})\right\rangle
\\&+0.006\left|^{4}T_{1g}(t_{2g}{}^{1}e_{g}{}^{2})\right\rangle -0.008\left|^{4}T_{1g}(t_{2g}{}^{2}e_{g}{}^{1})\right\rangle -0.002\left|^{4}T_{1g}(t_{2g}{}^{1}e_{g}{}^{2})\right\rangle \\&-0.266\left|^{2}T_{2g}(t_{2g}{}^{3})\right\rangle  +0.070\left|^{2}T_{2g}(t_{2g}{}^{2}e_{g}{}^{1})\right\rangle -0.014\left|^{2}T_{2g}(t_{2g}{}^{2}e_{g}{}^{1})\right\rangle \\&-0.011\left|^{2}T_{2g}(t_{2g}{}^{1}e_{g}{}^{2})\right\rangle    +0.006\left|^{2}T_{2g}(t_{2g}{}^{1}e_{g}{}^{2})\right\rangle -0.038\left|^{4}T_{2g}(t_{2g}{}^{2}e_{g}{}^{1})\right\rangle \\&-0.100\left|^{4}T_{2g}(t_{2g}{}^{2}e_{g}{}^{1})\right\rangle .
\end{split}
\end{equation}


\end{document}